\documentclass[12pt,a4paper]{article}
\usepackage{times}
\usepackage{durhampaper}
\usepackage{amsmath}
\usepackage{harvard}
\usepackage{graphicx}
\usepackage{xcolor}
\usepackage{url}
\usepackage{caption}
\usepackage{subcaption}
\usepackage{import}
\usepackage{bm}

\citationmode{abbr}
\title{IMU Tracking of Kinematic Chains in the Absence of Gravitational and Magnetic Fields}
\author{} 
\student{Greg K. Stretton}
\supervisor{Dr George A. Koulieris}
\degree{MEng Computer Science}

\date{May 2021}

\begin{document}

\maketitle

\begin{abstract}

{\bf Context/Background:} Tracking kinematic chains has many uses from healthcare to virtual reality. Inertial measurement units (IMUs) are well-recognised for their body tracking capabilities, however, existing solutions rely on gravity and often magnetic fields for drift correction. As humanity's presence in space increases, systems that don't rely on gravity or magnetism are required. {\bf Aims:} We aim to demonstrate the viability of IMU body tracking in a microgravity environment by showing that gravity and magnetism are not necessary for correcting gyroscope-based dead-reckoning drift. We aim to build and evaluate an end-to-end solution accomplishing this. {\bf Method:} A novel algorithm is developed that compensates for drift using local accelerations alone, without needing gravity or magnetism. Custom PCB sensor (IMU) nodes are created and combined into a body-sensor-network to implement the algorithm and the system is evaluated to determine its strengths and weaknesses. {\bf Results:} Dead-reckoning alone is accurate to within 1\textdegree{} for ~30s. The drift correction solution can correct large drifts in yaw within 4 seconds of lateral accelerations to within 3.3\textdegree{} RMSE. Correction accuracy when drift-free and under motion is 1.1 \textdegree{} RSME. {\bf Conclusions:} We demonstrate that gyroscopic drift can be compensated for in a kinematic chain by making use of local acceleration information and often-discarded centripetal and tangential acceleration information, even in the absence of gravitational and magnetic fields. Therefore, IMU body tracking is a viable technology for use in microgravity environments.
\end{abstract}


\section{Introduction}

    
    A kinematic chain is an assembly of rigid bodies connected by joints to provide constrained motion. These are everywhere in the modern world, from robotic arms to human bodies. It's often desirable to know information about the pose (position of limbs) of such systems, over time; application areas include healthcare, sports science, animation and virtual reality. We refer to all tracking systems that accomplish this, for artificial or biological systems, as ``body tracking systems" from now on.
    
    \subsection{Applications of body tracking}
    
    
    Motion tracking has many uses in the film industry as the data they provide are often key in special effects generation and animation. This often involves high-cost studio setups with green screens and expensive cameras, so low-cost motion tracking solutions can help make low-budget film production more viable.  Body tracking from a physiological perspective is important in healthcare and sports as it enables detailed analysis of movements, useful for optimising health and performance.
    
    
    As we increasingly inhabit digital spaces, body tracking for games and virtual reality is an especially relevant area. In the past visual methods have been used, like with the Kinect \cite{kinect}, but this base-station approach constrains the user to a small area where motion can be tracked. They also suffer occlusion and low-light issues. Modern VR like the Quest 2 \cite{quest2} tend to just track head and hands position. While this is sufficient for some uses, it places a cap on immersion. It's hard to feel immersed in a virtual space if your body is fundamentally disconnected from the experience. IMU body tracking solves these issues, enabling even multi-room VR experiences where motion can be tracked without a limit on the range, but such systems are not well adopted yet.

    \subsection{IMUs for body tracking}

    IMUs are devices with gyroscopes, accelerometers and sometimes magnetometers that can accurately estimate their orientation in the world over time. Therefore if one of these is placed on each limb of a human or robotic system, its pose and movements can be tracked over time. Gyroscopes drift over time, but gravitational and magnetic fields can provide fixed reference points to correct this. The cost to manufacture such devices has dropped over time and it's now possible to put one of these on every limb of a person for an accessible price \cite{xsens-mtw-awinda}. These systems can also allow for more flexibility in the design of robotic systems as they remove the need for mechanical trackers at joints. From industrial to military robotics, the trend is towards increasing speed and agility, so lightweight trackers like these could be attractive soon.
    
    IMU body tracking overcomes issues faced by optical methods like occlusion and lighting issues, making them flexible to be used in a wide range of settings. This work focuses on how drift can be corrected in IMU body tracking systems. Standard approaches rely on gravity and magnetism, a requirement that needs to be relaxed as we move forward.
    
    \subsection{Increasing human presence in space}
    
    There has now been a constant human presence in space for over twenty years at the International Space Station (ISS) \cite{iss}. This and the fact that over 3000 research investigations have been carried out in that time shows the strong demand for space-based research. Despite the post-Apollo regression, humanity is once again heading outwards with vigour, thanks to the increasing privatisation of the space industry.
    
    Motion tracking has many potential uses relating to space. Extravehicular activity (EVA) is an activity done outside a spacecraft, in a vacuum. This is a high-risk activity in a hazardous environment, so being able to track an astronaut's movements could provide critical information to ground control. Body tracking is already used on Earth for exercise and sports science \cite{IMU-exercise}, and it would be useful for this in space too - exercise is critical in microgravity due to muscle and bone-density loss. Body tracking systems in space could help people adjust to exercise in microgravity.
    
    As Earth-to-orbit launch costs continue to fall, we may soon see a rapid movement towards the industrialisation and commercial development of space. This will create requirements for body tracking capabilities in many novel environments. Aside from microgravity environments like the ISS, it's possible to substitute gravity for centrifugal force in future space habitats by spinning them along one axis. Perceived gravity in an O'Neill cylinder \cite{colonization-of-space} or Stanford torus \cite{space-settlements-study} would vary with radial movement, from full acceleration on the circumference to a zero-g environment at the centre. IMU body tracking systems for use in space should therefore be tolerant to arbitrary acceleration fields.
    
    Existing solutions for IMU body tracking rely on acceleration due to gravity for drift correction. Also, magnetometers are frequently used to correct yaw-drift, however, this is only possible due to Earth's strong magnetic field, such techniques would be of little help on other planets, moons, or in free space. A solution for IMU body tracking drift correction that relies on neither gravity nor magnetism is therefore needed as humanity increases its presence in space.
    
    \subsection{Contribution}
    
    Our proposed method for drift correction correlates rotational kinematics to linear accelerations to extract world frame accelerations of each pair of connected segments at their joint, revealing orientation information to be used for drift correction. 
    
    The drift correction method can be described simply in the following way: In a kinematic chain, with one sensor per limb, where all joints are fixed and the only accelerations are shared globally due to acceleration of the whole system, world frame acceleration vectors provide information about the joint angles. When the segments are moving there are various other accelerations, but these can be calculated and removed via circular motion equations based on the gyroscope values and kinematic chain information, and the aforementioned principle holds.
    
    This paper formulates a novel algorithm to achieve this, along with a hardware implementation demonstrating the viability of the approach. The implementation features custom sensor PCBs and a central hub to enable inter-sensor communication. The output data is available over Wi-Fi, making the body pose information easily accessible to software on the network, or even over the internet. A visualisation is developed in 3D graphics software to display the corrected orientations of each segment of a tracked kinematic chain.

\section{Related Work}

    \subsection{Representing rotation}
        There are many areas where it's necessary to mathematically represent rotations, from 3D graphics to spaceship flight software. It is also the most fundamental element of any IMU or body tracking system. Euler angles are commonly used as they are simple to understand, using 3 parameters of yaw, pitch, and roll \cite{rotation}. Although intuitive, they have significant problems. The biggest problem of gimble lock occurs when two of the axes align, resulting in the loss of a degree of freedom. Also, attempting to interpolate between two angles can give poor results if the axes are poorly aligned.
        
        The axis-angle representation where rotations are specified as some angle around some axis is also intuitive but overcomes the problems of Euler angles. Unit quaternions (a subset of the 4-dimensional complex numbers) are similar to axis-angles but are more powerful due to their properties. They can represent both orientations and rotations with perfect interpolation and no gimble lock. In this work we denote a quaternion $\boldsymbol{q}$ representing some rotation $\theta$ about some axis $\vec{v}$ as $\boldsymbol{q} = [\theta, \vec{v}]$, where $\boldsymbol{q}$ is normalised to unit length 1. They can be multiplied together, and the following equation rotates a quaternion $\boldsymbol{q}$ by another quaternion $\boldsymbol{r}$:
        \begin{equation}
            \boldsymbol{q}' = \boldsymbol{q}\boldsymbol{r}\boldsymbol{q}^{-1}
        \end{equation}
    
    \subsection{Inertial Measurement Units (IMUs)}
        
        An IMU is a device with inertial sensors on - typically a gyroscope to measure angular rate, and accelerometers to measure linear acceleration. Changes in orientation relative to some initial state can be calculated accurately in the short-term by integrating the gyroscope readings over time. However, absolute orientation can not be known this way, and gyroscopes suffer from random and systematic noise (bias) that will cause the readings to drift from their true value in the long-term (several seconds). The accelerometer and magnetometer give information that is reliable long-term and provides absolute orientation relative to the world coordinate frame. The vertical direction, inferred from acceleration due to gravity, can correct the world's vertical axis. Orientation around the vertical axis can be inferred from the magnetic field, assuming it always points north.
        
        Information from multiple sensors needs to be combined in the process of sensor fusion. The simplest fusion algorithm is the complementary filter \cite{compvkalman}, example below.
        
        \begin{equation}
            x = \alpha x + (1 - \alpha) c
        \end{equation}
        
        Here, $x$ represents short term information like gyroscope-inferred orientation and $c$ is long-term accurate orientation data from the accelerometer. The strength of correction can be adjusted with $\alpha$, typically $\alpha \geq 0.95$. This acts as a high-pass filter on the gyroscope and a low-pass filter on the accelerometer, retaining the short-term accuracy of the gyroscope while slowly nudging the value towards the long term reading. This damping is needed because the accelerometer has lots of high-frequency noise and linear accelerations that will distort the perceived gravity vector. There are other filters like Madgwick's algorithm \cite{madgwick} which perform very well under standard scenarios with gravity and/or magnetic fields. To extend functionality to new situations we use the straightforward complementary filter in this work.
        
        Existing work all relies on gravity to compensate drift \cite{survey-IMU}. Gravity is seldom absent in most environments so there hasn't been a drive to address this yet. Gravity-based systems would not be able to compensate drift in a zero-g environment, and would drift over time to nonsense readings. There is no clear solution to this with a single sensor.
        
        Magnetometer-free solutions have been explored already as magnetic fields are subject to variations and absence on earth, whether due to electrical interference or large metal structures nearby. Technologies such as ultrawideband (UWB) can be used to tether the system to the world frame \cite{uwb}, but this relies on base-stations and emitting EM signals which may not always be appropriate.
        
        Therefore, the solution presented in this work relies on neither gravity nor magnetism.
        
    \subsection{Multiple IMU systems}
    
        IMU body tracking is well researched and is increasingly being adopted recently due to reducing cost, increasing applications, and more companies being involved \cite{survey-IMU}. IMU systems have many benefits over optical alternatives, not suffering from occlusion, lighting issues, or location constraints.
        
        Commercial solutions for IMU human body tracking have existed for many years. Notably the \emph{Xsens MVN} \cite{xsens-mvn}, and the more recent \emph{Xsens MTw Awinda} \cite{xsens-mtw-awinda} which is available commercially. Xsens solutions use the accelerometers to "determine the direction of the local vertical by sensing acceleration due to gravity", and magnetometers to "[stabilise the] horizontal plane by sensing the direction of the earth magnetic field". Therefore the system could not function in a zero-g environment. Nevertheless, earth-based solutions are quite effective, with the Xsens specifying accuracy levels of 0.75\textdegree RMSE  for roll and pitch, and 1.5\textdegree RMSE for heading (yaw) with latency around $19ms$.
        
        Body tracking IMU systems are a subset of body sensor networks (BSNs). BSNs are well studied due to an increasing number of applications with the trend towards IoT \cite{BSNs}. While often completely wireless, inter-sensor communication could be wired to simplify design at a cost of reduced user convenience. For IMU systems it's now viable to create fully wireless sensors, as is done by the \emph{Xsens MTw Awinda} which achieves performance similar to wired, synchronising to within $10\mu s$, showing the viability of wireless for such systems. This requires a lot more design work and overhead though, so a wired solution is sufficient for prototyping.
        
        Much effort in the design of these systems goes to ensuring each sensor is rigidly bound to absolute references of the world frame. But perhaps this is not necessary. In the case where absolute orientations are required, this could be implemented on just the root of a kinematic chain like the head, with all child limbs being corrected relative to the root. And in cases like zero-g environments where the concept of absolute orientation is less well-defined, the system could be untethered - there is still plenty of utility in knowing the orientations of limbs relative to each other.
        
        It can be easier to measure limb angles directly with mechanical trackers for artificial systems like robotic arms. This can give high accuracy measurements of relative limb orientations \cite{survey-IMU}. These would add bulk and complexity to a system though, so IMUs could be considered as an alternative. IMUs are simple, light, solid-state devices and can be placed at any point on a segment, making them more flexible than joint-rotor tracking methods. 
    
    \subsection{Centripetal and tangential accelerations}
        Circular motion equations can be formulated to directly calculate accelerations on a rigid body with angular velocity and angular acceleration. This is highly relevant to IMU kinematic chain tracking - the gyro data can help offset unhelpful acceleration readings. In many systems though, this is not taken advantage of - centripetal and tangential accelerations are neglected due to being considered insignificantly small compared to gravitational and linear accelerations \cite{centri-discard}.
        
        Using this data to improve system accuracy has been investigated with promising results \cite{body-constraints}. The cited paper correlates between rotational kinematics and linear accelerations to correct the effects of linear accelerations on orientation predictions. This is useful when using gravity to represent the vertical axis, as linear accelerations and gravity get mixed up. This allowed for greater accuracy in orientation estimates. Their solution still relies on gravity as an absolute reference, presenting an opportunity to utilise similar principles to break free from this constraint.
        
        We will make use of the linear acceleration information provided by the angular rates, similar to \cite{body-constraints}, to estimate the acceleration vector at the base of a limb. This vector is then compared to the measurement from the parent limb, and drift is corrected based on the difference between these two vectors in the world frame. This method relaxes the requirement for a gravitational field.
        
\section{Solution}

        This solution tracks, and corrects drift in, the orientations of parts of a kinematic chain. This could be a robotic system or a human body. It's important to note that the global orientation of the tracked system can not be known, though internal limb orientations can be known relative to the root of the chain.

        The system assumes a kinematic chain such as the one shown in figure \ref{fig:kinematic_chain}. A sensor must be attached to each limb. This measures angular rate and acceleration. The sensors can communicate with each other and send output via a central hub which is available over WiFi.

        There are two elements to the algorithm: conventional orientation prediction by dead-reckoning and the novel drift correction method.
        
        Figure \ref{fig:chain_close} shows the detailed setup of a limb and its parent limb/sensor. The solution assumes the constraint that the sensor S is positioned exactly at the end of the limb and aligned to its coordinate axis. There is no distinction between sensor frame and limb frame. This assumption simplifies the maths and removes the need for calibration. The assumption is easily relaxed by adding in another frame's orientation data, but this is left for further work.
        
    \subsection{Algorithm}
    \subsubsection{Dead reckoning}
        Dead reckoning is used on each sensor $S_i$ to track its orientation by integrating the gyroscope readings. Gyroscope data $\omega_i = \{\omega_x, \omega_y, \omega_z\}$ is provided by the sensor in the body frame $B_i$. Updating the orientation $\boldsymbol{q}_i$ of the sensor is straightforward by composing a rotation quaternion $\boldsymbol{r}$ via an axis-angle representation of $\omega_i$. The updated orientation $\boldsymbol{q}_i'$ can then be found by rotating $\boldsymbol{q}_i$ by $\boldsymbol{r}$.
        
        \begin{equation}
            \begin{split}
                \boldsymbol{r} &= \Big[dt\Vert\omega_i\Vert, \frac{\omega_i}{\Vert\omega_i\Vert}\Big] \\
                \boldsymbol{q}_i' &= \boldsymbol{r}\boldsymbol{q}_i\boldsymbol{r}^{-1}
            \end{split}
        \end{equation}
            
        This is the core of the algorithm and is accurate in the short term (several seconds). Dead-reckoning accumulates errors over time, hence the need for drift correction, described in \ref{correction}. The high short-term accuracy of the above is preserved via a complementary filter that only partially accepts the drift correction result.
        
    \subsubsection{Calculating drift} \label{calculate-drift}
            
        \begin{figure}[h!]
            \centering
            \begin{subfigure}{.45\textwidth}
                \centering
                \vspace{1.5cm}
                \includegraphics[width=0.9\linewidth]{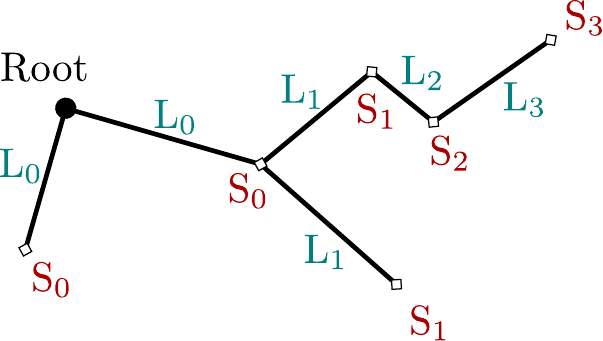}
                \vspace{1.5cm}
                \caption{An example kinematic chain comprising limbs $L$ with sensors $S$ attached at each limb's tip.}
                \label{fig:kinematic_chain}
            \end{subfigure}\qquad
            \begin{subfigure}{.45\textwidth}
                \centering
                \includegraphics[width=0.9\linewidth]{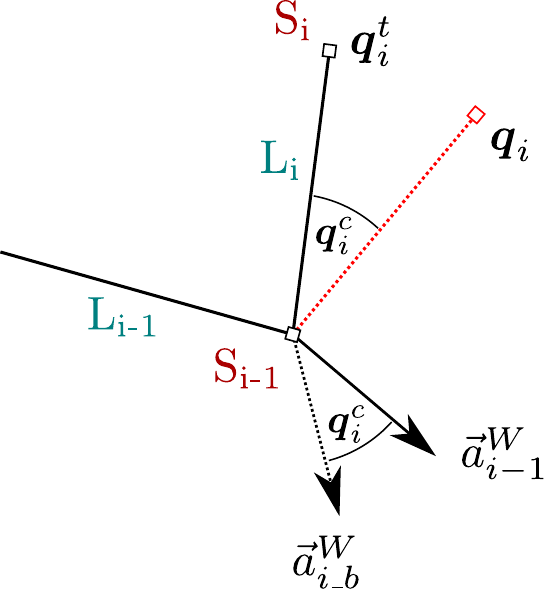}
                \caption{Parent/child ($L_{i-1}$/$L_i$) joint setup. Showing the correction quaternion $\boldsymbol{q}^c_i$, calculated from the global acceleration vectors of each limb at the joint.}
                \label{fig:chain_close}
            \end{subfigure}
            \caption{Kinematic chain setup.}
            \label{fig:board}
        \end{figure}
        
        Consider a setup with two limbs, child $L_i$ and parent $L_{i-1}$, with the base $L_i^b$ of the child limb joined to the tip $L_{i-1}^t$ of the parent as shown in figure \ref{fig:chain_close}. We assume the orientation of $L_{i-1}$ is known to be $\boldsymbol{q}_{i-1}$. The orientation $\boldsymbol{q}_i$ of $L_i$ stored by the system has drifted from its true value $\boldsymbol{q}_i^t$ by some angle $\boldsymbol{q}_i^c$. This stage of the algorithm calculates $\boldsymbol{q}_i^c$ and later uses this to correct the drift.
        
        For every limb $L$, a sensor $S$ is placed at the tip $L^t$. Gyroscope values $\omega$ and the acceleration vector $\vec{a}$ are available from the sensor. These readings are all relative to the sensor's body frame $B$. A vector $\vec{a}$ can be transformed to the world frame $W$ by rotation of $\boldsymbol{q}$, giving $\vec{a}^W$; this is an important operation.
            
        $L_i^b$ and $L_{i-1}^t$ are at the same point in space, so the acceleration vectors $\vec{a}_{i\_b}$ and $\vec{a}_{i-1}$ must be equal in $W$, where $\vec{a}_{i\_b}$ is the acceleration in $B_i$ at $L_i^b$. $\vec{a}_{i\_b}$ can be calculated by removing centripetal and tangential acceleration from $\vec{a}_i$, shown in A.3. Since these two acceleration vectors are identical, ignoring noise, any difference between them in the world frame is due to error in $B_i \rightarrow W$, and therefore in $\boldsymbol{q}_i$. The rotation from $\vec{a}_{i\_b}^W$ to $\vec{a}_{i-1}^W$ is therefore $\boldsymbol{q}_i^c$. It is straightforward to compose the quaternion $\boldsymbol{q}_i^c$ from an axis-angle representation $[w, xyz]$:
        
        \begin{equation}
            \label{correction}
            \begin{split}
                \vec{a}_{i-1}^W &= \boldsymbol{q}_{i-1} \vec{a}_{i-1} \boldsymbol{q}_{i-1}^{-1}\\
                \vec{a}_{i\_b}^W &= \boldsymbol{q}_i \vec{a}_{i\_b} \boldsymbol{q}_i^{-1} \\
                \theta &= \arccos \Big( \frac{\vec{a}_{i-1}^W \cdot \vec{a}_{i\_b}^W}{\Vert\vec{a}_{i-1}^W\Vert \cdot \Vert\vec{a}_{i\_b}^W\Vert}\Big)\\
                xyz &= \vec{a}_{i-1}^W \times \vec{a}_{i\_b}^W \\
                \boldsymbol{q}_i^c &= [\theta, xyz] \\
            \end{split}
        \end{equation}
        
    \subsubsection{Calculating acceleration due to rotation.}
        \label{accels}
    
        \begin{figure}[htb!]
            \centering
            
            \begin{subfigure}{.45\textwidth}
                \centering
                \includegraphics[scale=1.3]{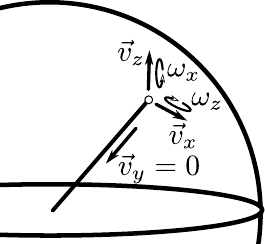}
                \caption{Calculating velocity from angular rates.}
                \label{fig:circ_vel}
            \end{subfigure}
            \begin{subfigure}{.45\textwidth}
                \centering
                \includegraphics[scale=0.5]{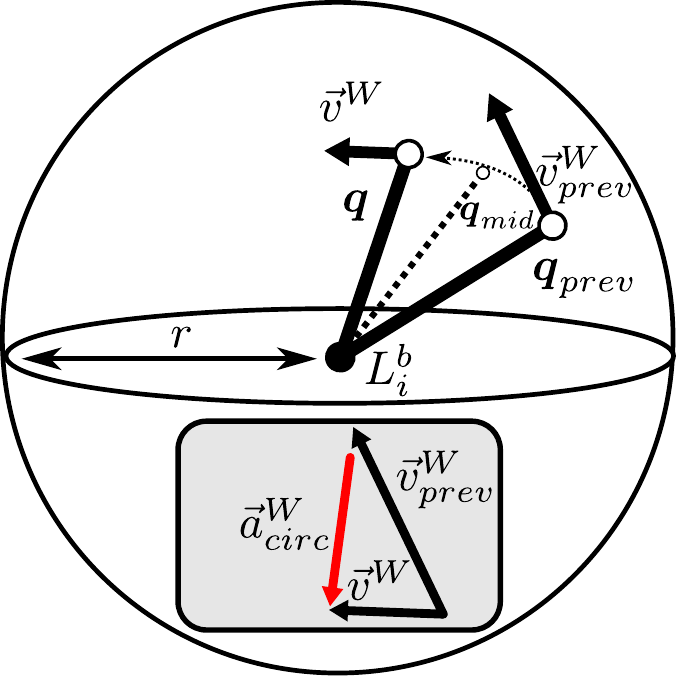}
                \caption{Acceleration due to circular motion.}
                \label{fig:circ_motion}
            \end{subfigure}%
            \label{fig:circ}
            \caption{Circular motion setup.}
        \end{figure}
        
        A limb $L_i$ is assumed to be of fixed length $r_i$, so the tip $L^t_i$ can be modelled as moving on the surface of a sphere of radius $r_i$ with centre $L^b_i$. Therefore, the acceleration at the base $\vec{a}_{i\_b}$ is the acceleration at the tip $\vec{a}_i$ minus any accelerations $\vec{a}_{circ}$ caused due to angular rates about the sphere.
        
        The linear velocity of the tip can be calculated from the angular rate of the perpendicular gyroscope reading. As shown in figure \ref{fig:circ_vel}, rates of $\omega_x$ correspond to $\vec{v}_z$, and $\omega_z$ to $\vec{v}_x$, by the equation $\vec{v}=r\omega$. As $r_i$ is fixed, $\vec{v}_y$ is always 0. The current and previous-update velocities can be found in this way and transformed from the body to the world frame for meaningful comparison:

        \begin{equation}
        \label{acceq1}
        \begin{split}
            \vec{v}_{prev} &= [r\omega_{z\_prev}, 0, r\omega_{x\_prev}]\\
            \vec{v} &= [r\omega_z, 0, r\omega_x]\\
            \vec{v}_{prev}^W &= \boldsymbol{q}_{prev}\vec{v}_{prev}{\boldsymbol{q}_{prev}}^{-1}\\
            \vec{v}^W &= \boldsymbol{q}\vec{v}\boldsymbol{q}^{-1}\\
        \end{split}
        \end{equation}
        
        Acceleration can now be calculated from the change in velocity found in equation \ref{acceq1}. The result is then transformed back into the body frame for subtraction from the sensor reading $\vec{a}_i$. Since predictions are most effective under rapid motion (shown in IV.\ref{accelpredresults}) we add a weighting function $\beta$ that scales contribution based on angular rate and angular acceleration.
        
        \begin{equation}
        \label{acceq2}
        \begin{split}
            \vec{a}_{circ}^W &= \frac{\vec{v}^W - \vec{v}_{prev}^W}{dt} \\
            \vec{a}_{i\_b} &= \vec{a}_i - \beta({\boldsymbol{q}_{mid}}^{-1}\vec{a}_{circ}^W\boldsymbol{q}_{mid})\\
        \end{split}
        \end{equation}
        
        $\vec{a}_{i\_b}$ is then fed into equation \ref{correction}. When combined, the equations can be simplified to use fewer transformations, but they are presented in this way here for clarity.
            
    \subsubsection{Correcting drift - Complementary filter}\label{compfilter}
            
        The above assumes a noiseless environment, but in reality accelerometer readings $\vec{a}$ suffer from noise. Therefore, any reading $\vec{a}$ is the sum of signal vector $\vec{a}_s$ and noise vector $\vec{a}_n$. This means that $\boldsymbol{q}_i^c$ is unreliable for small values of $\Vert\vec{a}_{i-1}^W\Vert\cdot\Vert\vec{a}_{i\_b}^W\Vert$. The mean noise magnitude $\mu(\Vert\vec{a}_n\Vert)$ is determined in part \ref{sensor-noise-results} of results. The signal-to-noise ratio SNR can be calculated as follows:

        \begin{equation}
            \begin{split}
                \text{SNR} &= \frac{\Vert\vec{a}_{i-1}^W\Vert\cdot\Vert\vec{a}_{i\_b}^W\Vert}{\mu(\Vert\vec{a}_n\Vert)^2}\\
            \end{split}
        \end{equation}
        
        Instead of applying the full $\boldsymbol{q}_i^c$, we construct a complementary filter to apply the drift correction to $\boldsymbol{q}_i$ with varying intensity based on the SNR. The SNR is mapped to range $(0,1)$ by the function $\Gamma$:
        
        \begin{equation}
            \begin{split}
                \phi &= 0^c\cdot(1 - \Gamma(\text{SNR})) + \theta\cdot\Gamma(\text{SNR})\\
                \boldsymbol{q}_i^\Gamma &= [\theta\cdot\Gamma(\text{SNR}), xyz]\\
            \end{split}
        \end{equation}
        
        There are several possibilities for $\Gamma$, here we choose a simple linearly increasing function.
            
        The corrected orientation $\boldsymbol{q}_i'$ can now be found by rotating $\boldsymbol{q}_i$ by $\boldsymbol{q}_i^\Gamma$:
        
        \begin{equation}
            \boldsymbol{q}_i' = \boldsymbol{q}_i^\Gamma\boldsymbol{q}_i(\boldsymbol{q}_i^\Gamma)^{-1}
        \end{equation}

    \subsection{Sensor node}
\begin{figure}[h!]
            \centering
            \begin{subfigure}{.5\textwidth}
                \centering
                \includegraphics[width=.9\linewidth]{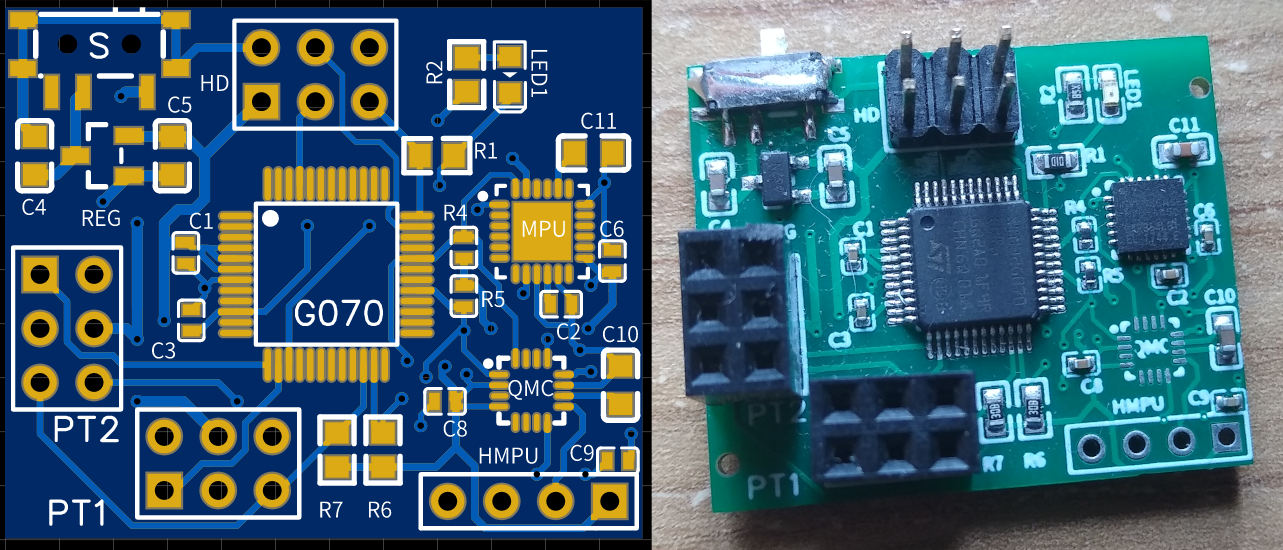}
                \caption{The PCB design and an assembled sensor node.}
                \label{fig:board_image}
            \end{subfigure}%
            \begin{subfigure}{.5\textwidth}
                \centering
                \includegraphics[width=.9\linewidth]{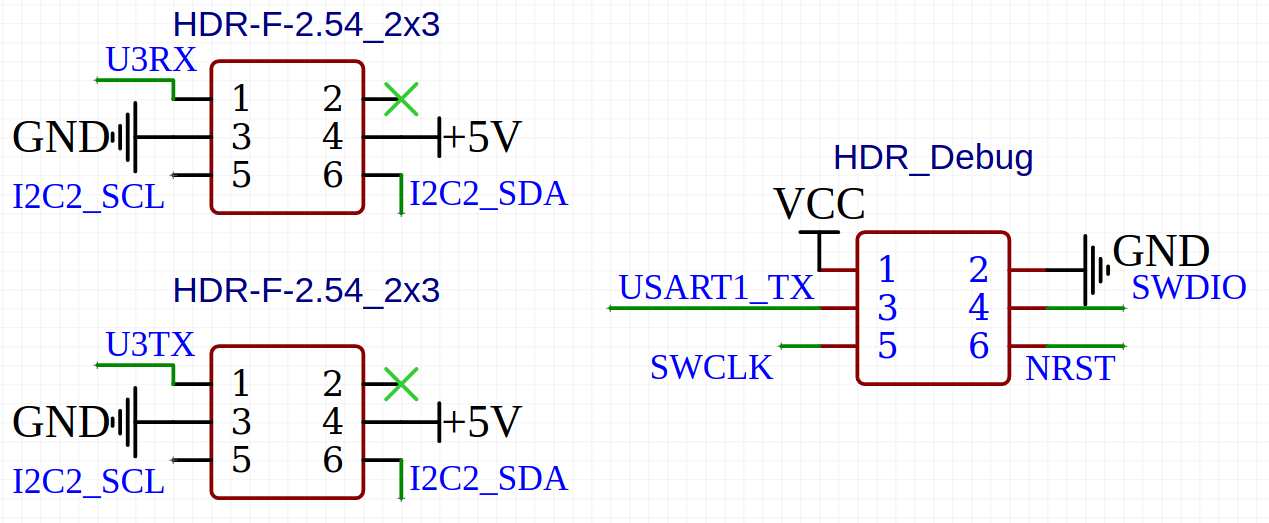}
                \caption{Communication and debug pin-out.}
                \label{fig:pinout}
            \end{subfigure}
            \caption{Board images}
            \label{fig:board}
        \end{figure}

        This section outlines the design of the individual sensor nodes. Most of the processing is done by the sensor nodes, with the central hub just relaying information. Figure \ref{fig:board_image} shows the PCB design and assembled node. Here are the major design requirements:
        
        \begin{itemize}
            \item High-performance processor to implement the algorithm at a maximal update rate.
            \item Able to get accelerometer and gyroscope sensor data at a high rate (100Hz+). 
            \item A sensor can send its acceleration data to its child sensor(s) and each sensor can send its orientation to a central hub (sensor-hub communication).
        \end{itemize}
        
        An STM32G070CBT6 \cite{stm} microprocessor is on the board. This has up to 64MHz with plenty of memory and sufficient peripherals for the implementation. The MPU-6050's \cite{mpu} gyroscope and accelerometer were included to meet the sensor data requirement. Various other electronic components were required, including a voltage regulator, decoupling capacitors, and I\textsuperscript{2}C pull-up resistors. A status LED and custom debug header interface for in-system programming were also added to ease development.
        
        I\textsuperscript{2}C was chosen as the sensor-hub communication protocol. Although it was designed for use on a single board, longer distances can be supported by strengthening the pull-up resistors. SPI was considered as an alternative due to its higher speed, but the extra 2 wires would have made the system less convenient, and I\textsuperscript{2}C in fast-mode cycles at 400KHz which is sufficient. The two 2x3 female headers are the communication interfaces; their pin-out is shown in figure \ref{fig:pinout}.
        
        Some redundancy was included in the design to offset the risk of part of the system not working: A UART interface was added in case the I\textsuperscript{2}C interface had problems over the long distances and an interface for an MPU-9250 module board was added as a sensor backup. High-current lines were kept away from the sensors to prevent interference.
        
        A QMC5883L \cite{qmc} magnetometer was designed in but not included on the final board assembly due to a manufacturer supply issue. This apparent inconvenience inspired the magnetometer-free implementation and space application ideas. The boards were designed using \url{EasyEDA.com} and ordered from \url{JLCPCB.com}.
    
    \subsection{Sensor network}
\begin{figure}[!htb]
        \centering
        \begin{subfigure}{0.45\linewidth}
            \centering
            \includegraphics[width=\linewidth]{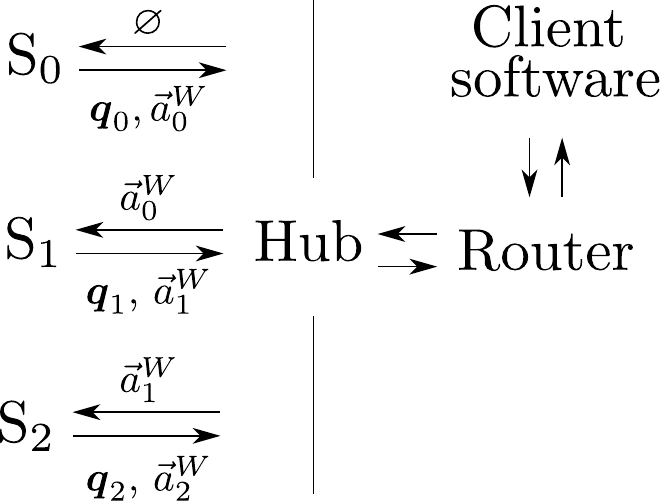}
            \caption{System architecture and communication structure between hub and sensors.}
            \label{fig:architecture}
        \end{subfigure}\qquad
        \begin{subfigure}{0.45\linewidth}
            \centering
            \includegraphics[width=\linewidth]{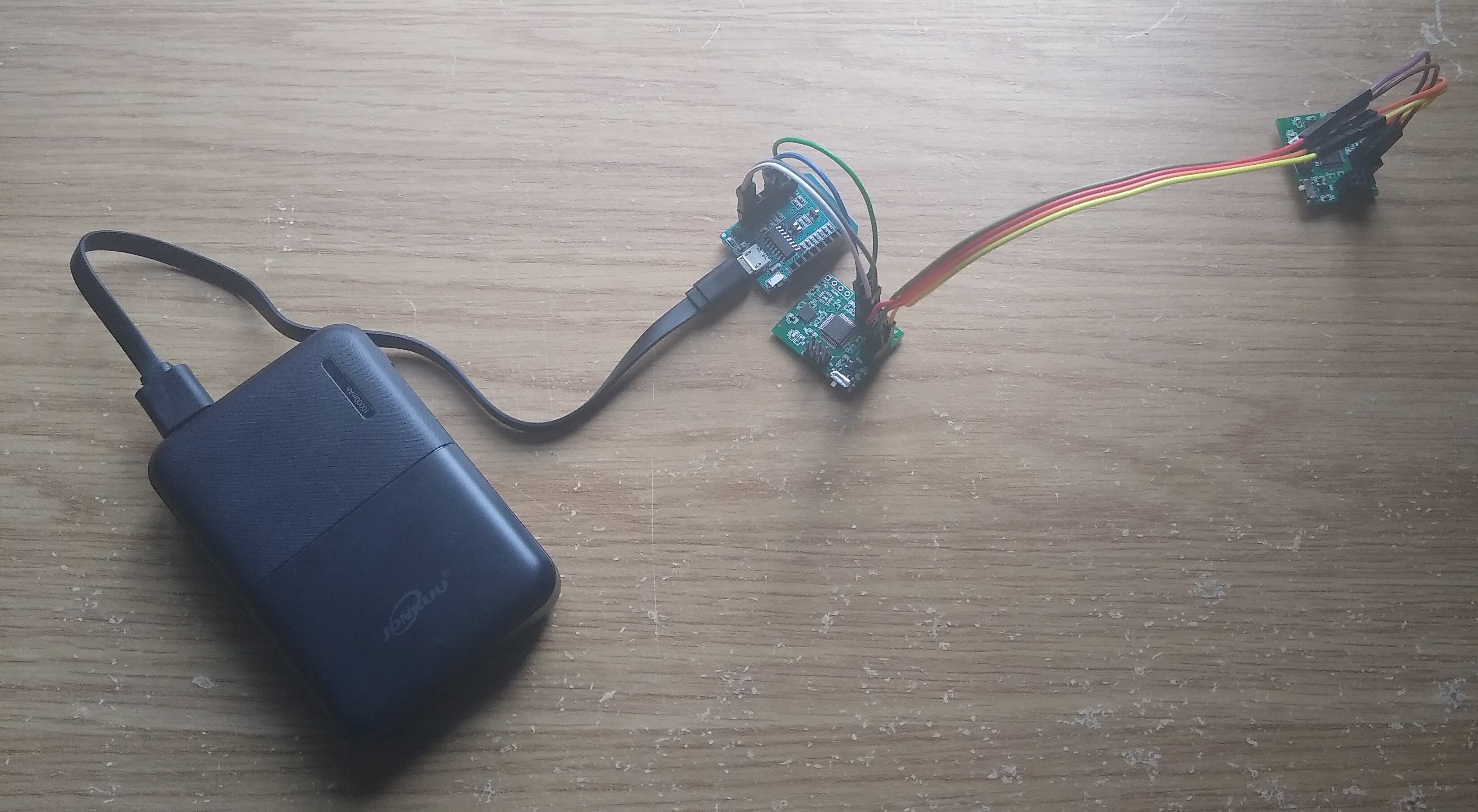}
            \caption{An example setup with two sensors, not attached to a kinematic chain.}
            \label{fig:physical}
        \end{subfigure}
        \end{figure}

        An architecture is required to coordinate inter-sensor communication and make the sensor orientations available to any client software. The architecture should minimise latency and maximise bandwidth to ensure accuracy and scalability. The high-level architecture is shown in figure \ref{fig:architecture}.
        
        The sensors send their orientations over I\textsuperscript{2}C to the ESP8266 Wi-Fi server, which then makes the data available in JSON format via an HTTP request. Any client program can therefore read the data off over a network router. The I\textsuperscript{2}C traffic pattern is shown in figure \ref{fig:architecture}.
        
        The hub, being the I\textsuperscript{2}C master, initiates and coordinates drift correction. Each sensor is sent to, then received from, in order, looping back to the start once finished. This is a simplification considering just one branch of the kinematic chain - the full implementation would be a depth-first traversal of the full tree e.g. figure \ref{fig:kinematic_chain}. When a sensor $S_i$ receives its parent's acceleration data from the hub (or dummy data for $S_0$), it calculates and corrects for drift in the system. $S_i$ then sends its orientation $\boldsymbol{q}_i$ and acceleration $\vec{a}_i^W$ to the hub. The hub makes $\boldsymbol{q}_i$, being the system's output, available over Wi-Fi, and $\vec{a}_i^W$ is sent to $S_{i+1}$ to initiate the next step.
        
        This communication is accomplished over I\textsuperscript{2}C, with the hub being the master and each sensor an I\textsuperscript{2}C slave. Bandwidth is limited here, so the data format is important. Values of $\boldsymbol{q}$ and  $\vec{a}$ are arrays of doubles. This level of precision is unnecessary in inter-sensor transmission. Representing the values with shorts (2 bytes) instead of doubles (8 bytes) cuts bandwidth requirement by $4$. $\boldsymbol{q}$ and $\vec{a}$ have 4 and 3 values respectively. Therefore, 6 bytes must be sent to, and 14 from, each sensor in one communication loop. Bandwidth and latency for the I\textsuperscript{2}C and Wi-Fi networks are investigated in section \ref{network-analysis} of the results.
        
        Due to this architecture, drift correction is initiated on any given sensor infrequently from its point of view. Therefore, dead-reckoning runs at the maximal rate on the sensor and drift correction occurs by an interrupt from the hub. These timings are analysed in the results.
        
        Physically, the I\textsuperscript{2}C bus comprises 4 wires (+5V, GND, SDA, SCL). The system is powered with a rechargeable power pack, making the system wireless (in the sense of no fixed tethers). An example setup is shown in figure \ref{fig:physical} with two sensors that aren't attached to a kinematic chain.
       
    \subsection{Client program and physical setup}
     
        \begin{figure}[h!]
            \centering
            \begin{subfigure}{.45\textwidth}
                \includegraphics[width=\linewidth]{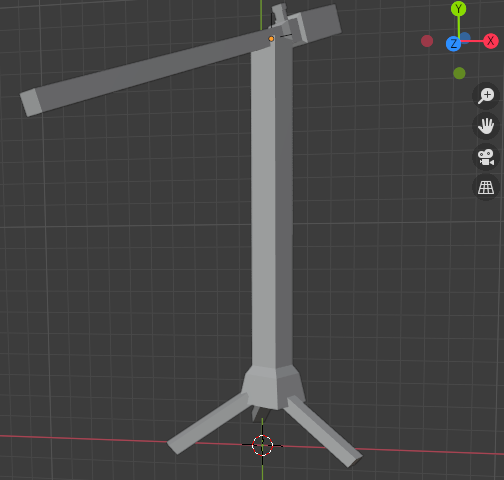}
                \caption{Blender visualisation}
                \label{fig:blender}
            \end{subfigure}\qquad
            \begin{subfigure}{.45\textwidth}
                \includegraphics[width=\linewidth]{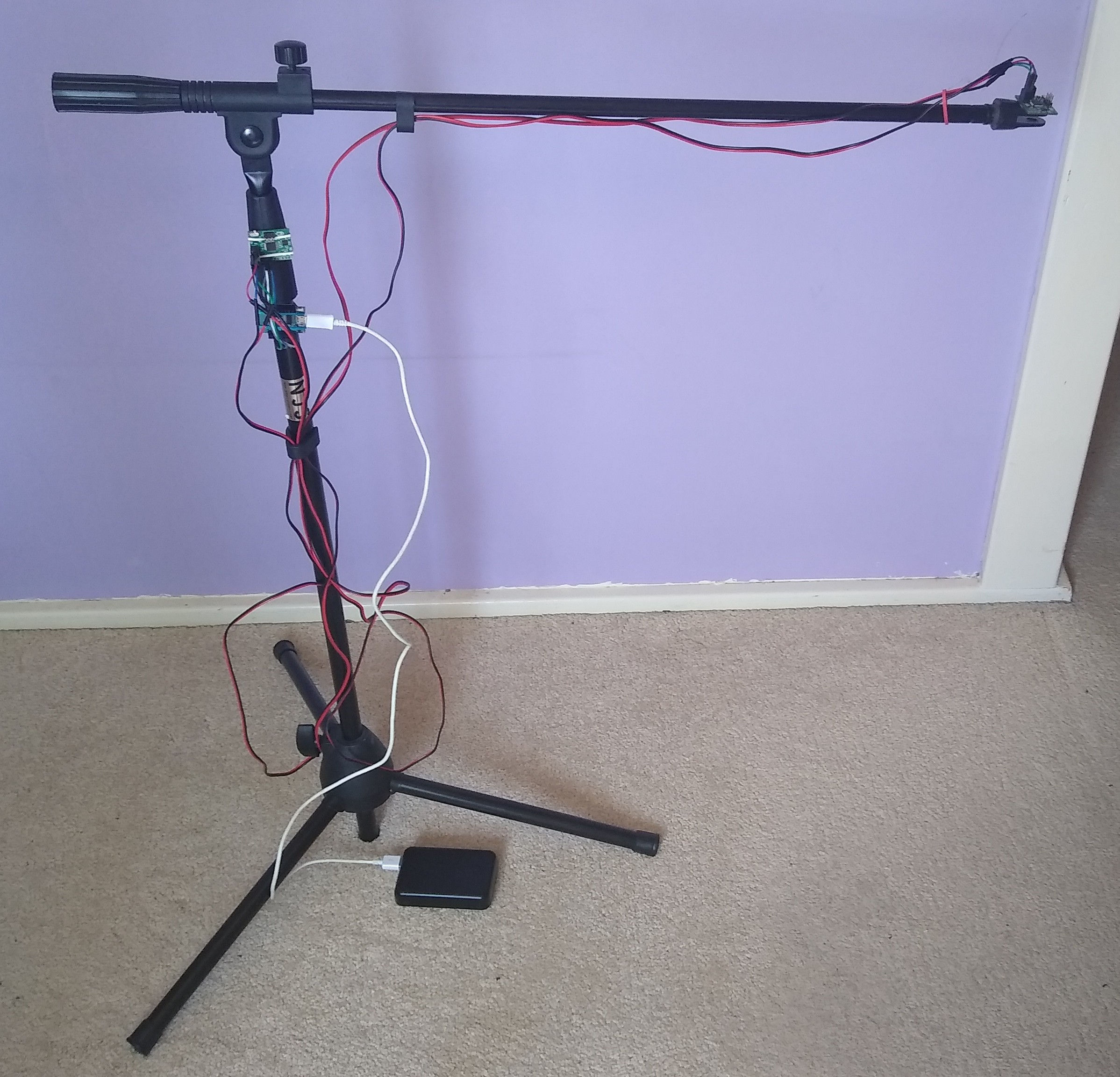}
                \caption{2-segment boom.}
                \label{fig:boom}
            \end{subfigure}
            \caption{Program and setup.}
            \label{fig:setups}
        \end{figure}
        
        The output data from the system could be used by any software, it is just JSON received by an HTTP request. For this work, a visualisation was built in the 3D graphics program Blender \cite{blender}. Blender is a very flexible program that supports Python scripting, making the visualisation setup straightforward. Several visualisations were made for demonstration purposes of different processes in the system, and the main one can be seen in figure \ref{fig:blender}. This detects the setup (3-segment arm vs 2-segment boom) and displays the output data in real-time with no noticeable latency.
        
        The main physical setup used is the 2-segment boom arm shown in figure \ref{fig:boom}. This was ideal for testing and result taking for several reasons. Firstly, the long arm increases accelerations due to angular rates, making it easier to analyse the performance of the prediction equations. Also, it is much easier to mount the sensors to a well-structured artificial system than a human body. Muscles move at different rates to the underlying bones, so fixing a coordinate frame is hard on organic systems. Furthermore, testing is a lot more convenient this way because the system can be left taking data for a long time.

\section{Results} \label{results}
    
    Details of results for the solution are given in the section, making use of a variety of metrics. The 2-sensor boom arm setup was used for all results. RMSE denotes the ``root mean square error" which measures the deviation of some data from target data. MAE, mean average error, is similar but weights all values equally, whereas RMSE gives more weight to more significant errors. The difference between these values gives an idea of the variance in error sizes, so they go well together. These can both be thought of as the average error size.
        
    \subsection{Sensor noise analysis} \label{sensor-noise-results}
     
        The baseline sensor noise is important to quantify as it gets passed into all stages of the algorithm. It is also needed by the complementary filter, as described in III.\ref{compfilter}. Below, the metrics outlined above are shown for the accelerometer and gyroscope.
           
        \begin{table}[!htb]
        \centering
        \caption{Base sensor noise of accelerometer and gyroscope.}
        \label{noise}
        \begin{tabular}{|l|l|l|} \hline
        \textbf{Reading} & \textbf{RMSE} & \textbf{MAE} \\ \hline
        \(\Vert\vec{a}\Vert\) (m\(\cdot\text{s}^{-2}\)) & 0.043 & 0.035  \\ \hline
        \(\Vert\vec{\omega}\Vert\) (rad\(\cdot \text{s}^{-1})\) & 0.0027 & 0.0025 \\ \hline
        \end{tabular}
        \end{table}

    \subsection{Dead reckoning}
    
        The need for drift correction comes from dead reckoning accumulating error over time. Here we quantify that drift, showing the need for correction and estimating how long readings remain accurate in its absence. Tables \ref{driftseconds} and \ref{driftangles} show that measurements remain accurate within 2 degrees for up to a minute. Measurements were taken just after calibration and in a stationary scenario, so these could worsen with more time and motion, though in general these results are highly accurate in the short term.
        
        \begin{table}[!htb]
        \begin{minipage}{.5\linewidth}
            \centering
            \caption{Drift times in seconds, n=56.}
            \label{driftseconds}
            \begin{tabular}{l|l|l|l|}
            \cline{2-4}
                & \multicolumn{3}{c|}{\textbf{Time to drift by x\textdegree}}                  \\ \hline
            \multicolumn{1}{|l|}{\textbf{Measure}} & \textbf{0.25\textdegree} & \textbf{0.5\textdegree} & \textbf{1.0\textdegree} \\ \hline
            \multicolumn{1}{|l|}{Mean (s)} & 7.4 & 14.8 & 29.7 \\ \hline
            \multicolumn{1}{|l|}{Maximum (s)} & 9.1 & 17.3 & 34.2 \\ \hline
            \multicolumn{1}{|l|}{Minimum (s)} & 5.6 & 12.1 & 26.6 \\ \hline
            \end{tabular}
        \end{minipage}
        \begin{minipage}{.5\linewidth}
            \centering
            \caption{Drift angles in degrees, n=28.}
            \label{driftangles}
            \begin{tabular}{l|l|l|l|}
            \cline{2-4}
                                                   & \multicolumn{3}{c|}{\textbf{Drift after x seconds}} \\ \hline
            \multicolumn{1}{|l|}{\textbf{Measure}} & \textbf{5s}     & \textbf{20s}    & \textbf{60s}    \\ \hline
            \multicolumn{1}{|l|}{Mean (\textdegree) }        & 0.15 & 0.59& 1.75\\ \hline
            \multicolumn{1}{|l|}{Maximum (\textdegree)}          & 0.19& 0.73& 2.13\\ \hline
            \multicolumn{1}{|l|}{Minimum (\textdegree)}          & 0.10& 0.40& 1.26\\ \hline
            \end{tabular}
        \end{minipage}
        \end{table}

    \subsection{Acceleration prediction}
        \label{accelpredresults}
        Here we show results for the accuracy of the correlation predictions between rotation and linear acceleration described in III.\ref{accels} and \cite{body-constraints}. The measures in table \ref{localnoise} show prediction noise when stationary. The y axis has far less noise as this is just the centripetal acceleration, whereas the other axes require differentiating the gyroscope to calculate tangential accelerations.
        
        \begin{table}[!htb]
        \centering
        \caption{Local acceleration prediction noise.}
        \vspace*{6pt}
        \label{localnoise}
        \begin{tabular}{|c|c|c|}\hline
        \textbf{Axis} & \textbf{RMSE} & \textbf{MAE} \\ \hline
        x (m\(\cdot\text{s}^{-2})\)& 0.17 & 0.13 \\ \hline
        y (m\(\cdot\text{s}^{-2})\)& 6.3e-6 & 4.5e-6 \\ \hline
        z (m\(\cdot\text{s}^{-2})\)& 0.14 & 0.11 \\ \hline
        \end{tabular}
        \end{table}
        
        To analyse these under motion, we transform to the world frame with \(\boldsymbol{q}\). Therefore, the vertical gravity vector is the target data in the table \ref{worldaccels} analysis. Data from each axis was combined, averaging out differences across axes. ``Slow movement" averaged 0.72 rad\(\cdot \text{s}^{-1}\) with peaks of 1.2 rad\(\cdot \text{s}^{-1}\) and ``fast movement'' averaged 2.13 rad\(\cdot \text{s}^{-1}\) with peaks of 3.9 rad\(\cdot \text{s}^{-1}\). Predicting acceleration reduces performance by a factor of 3 when still. However, prediction becomes increasingly effective with faster movements, being 2.7x and 4.9x more effective with slow and fast movements, respectively. This provides justification for weighting the prediction contribution by the angular rate/acceleration as described in III.\ref{accels}.
            
        \begin{table}[!htb]
        \centering
        \caption{World acceleration prediction accuracy (m\(\cdot\text{s}^{-2})\).}
        \label{worldaccels}
        \begin{tabular}{|l|l|l|l|l|l|l|}
        \hline
        \textbf{Scenario}  & \multicolumn{2}{l|}{\textbf{Stationary}} & \multicolumn{2}{l|}{\textbf{Slow movement}} & \multicolumn{2}{l|}{\textbf{Fast movement}} \\ \hline
        \textbf{Measure}   & \textbf{RMSE}        & \textbf{MAE}       & \textbf{RMSE}         & \textbf{MAE}         & \textbf{RMSE}         & \textbf{MAE}         \\ \hline
        No prediction & 0.037               &  0.029             & 1.22                 & 0.87                 & 6.03                 &  4.77                   \\ \hline
        With prediction    & 0.105               &  0.077             & 0.46                 & 0.31                 & 1.24                 &  0.91                 \\ \hline
        \end{tabular}
        \end{table}

    \subsection{Drift correction}
        Taking good results for the drift correction method within a strong gravitational field like Earth's is challenging. Note that this system is designed with a zero-g environment in mind, and the best experiments that would demonstrate the viability of this system are not possible on Earth. Table \ref{correctionaccuracy} evaluates the accuracy of the drift correction method. The system was initialised free from drift, and since dead-reckoning is accurate in the short term, \(\boldsymbol{q}\) is used as the ground truth orientation. The noise in the raw correction angle $\theta$ is smoothed out by the complementary filter, giving a much more stable final value of $\phi$. ``Moving" results averaged 1 rad\(\cdot \text{s}^{-1}\).
        
        \begin{table}[!htb]
        \centering
        \caption{Correction accuracy with and without complementary filter. (\textdegree)}
        \label{correctionaccuracy}
        \begin{tabular}{c|c|c|c|c|}
        \cline{2-5}
                                                        & \multicolumn{2}{c|}{\textbf{Stationary}} & \multicolumn{2}{c|}{\textbf{Moving}} \\ \hline
        \multicolumn{1}{|c|}{\textbf{Correction angle}} & \textbf{RMSE}        & \textbf{MAE}       & \textbf{RMSE}      & \textbf{MAE}     \\ \hline
        \multicolumn{1}{|c|}{$\theta$ (raw)}            & 3.5                 & 2.8                & 11.0              & 6.0              \\ \hline
        \multicolumn{1}{|c|}{$\phi$ (filtered)}         & 0.34                & 0.28               & 1.1               & 0.60             \\ \hline
        \end{tabular}
        \end{table}
        
\begin{figure}[!htb]
            \centering
            \includegraphics[scale=0.5]{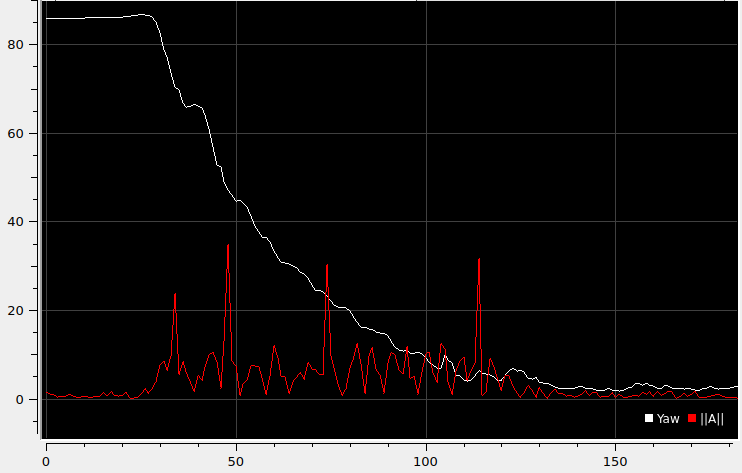}
            \caption{Live drift correction in yaw due to lateral accelerations. $\Vert \vec{a}\Vert$ denotes the magnitude of the lateral accelerations. The x-axis quantity is the update number ($30Hz$).}
            \label{fig:correctionexample}
        \end{figure}

        Figure \ref{fig:correctionexample} shows how this method out-performs conventional IMU tracking approaches. Here the system is initialised with a yaw drift of 90\textdegree. As with standard IMU approaches, this can not be corrected when still because the gravity vector gives no information of yaw. The figure shows the yaw correct from 90\textdegree{} to its true value of zero in 4 seconds under lateral accelerations in the horizontal plane.
        
        The above experiment was repeated 20 times and the final yaw measured to estimate the accuracy of correcting a large drift in a short time. On average, $\Vert\vec{a}_{lateral}\Vert$  was 7ms-2 over 4 seconds. After correction, the yaw had RMSE of 3.3\textdegree{} and MAE of 3.1\textdegree{}.
    
    \subsection{Network throughput/latency} \label{network-analysis}
        Latency is important for any IMU body tracking system. Measurements should be as up-to-date as possible to display potentially rapid body movements in real time. The update rate must also be high to prevent stuttering and maximise useful information. The I\textsuperscript{2}C response times and predicted operating rates can be seen below.
        
            \begin{table}[!htb]
            \begin{minipage}{.45\linewidth}
                \centering
                \caption{I\textsuperscript{2}C timings, subject to variation depending on the interrupt time.}
                \label{i2cresults}
                \begin{tabular}{c|c|c|c|}
                \cline{2-4}
                                                      & \multicolumn{3}{c|}{\textbf{Time $\mu s$}} \\ \hline
                \multicolumn{1}{|c|}{\textbf{Response time}} & \textbf{Min}  & \textbf{Mean}  & \textbf{Max} \\ \hline
                \multicolumn{1}{|c|}{Root sensor}    & 0             & 1150           & 2300         \\ \hline
                \multicolumn{1}{|c|}{Child sensor}   & 4100          & 5250           & 6400         \\ \hline
                \end{tabular}
            \end{minipage}\quad
            \begin{minipage}{.55\linewidth}
                \centering
                \caption{Operating rates, extrapolated from table \ref{i2cresults}.}
                \label{correctionrate}
                \begin{tabular}{|c|c|c|}
                \hline
                \textbf{Sensor quantity} & \textbf{Duration $\mu s$} & \textbf{Frequency $Hz$} \\ \hline
                2 (Test setup)           & 6400                    &  156                   \\ \hline
                3 (Single arm)           & 11650                          & 85                        \\ \hline
                7 (Upper body)           & 32650                          & 30                       \\ \hline
                15 (Full body)           & 74650                          & 13                       \\ \hline
                \end{tabular}
            \end{minipage}
            \end{table}

\begin{figure}[h!]
                \centering
                \begin{subfigure}{.5\textwidth}
                    \centering
                    \includegraphics[width=.9\linewidth]{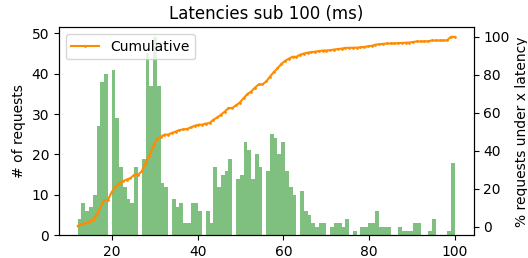}
                    \caption{Latency on BT Smart Hub 6.}
                    \label{fig:latencybt}
                \end{subfigure}%
                \begin{subfigure}{.5\textwidth}
                    \centering
                    \includegraphics[width=.9\linewidth]{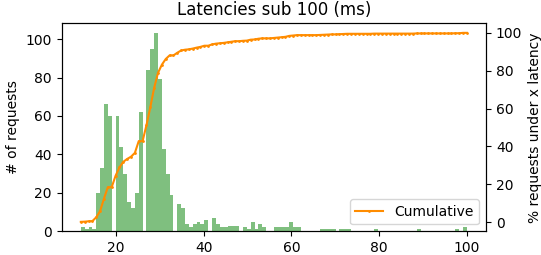}
                    \caption{Latency on tp-link TL-WR841N.}
                    \label{fig:latencytp}
                \end{subfigure}
                \caption{Wi-Fi response latencies with two sensors.}
                \label{fig:latencies}
            \end{figure}

            To determine the effect of network hardware on the Wi-Fi latencies, results were taken using two different routers. The charts in figure \ref{fig:latencies} show the number of packets against their latency. Figure \ref{fig:latencybt} uses a home router with other devices on the network. This has much less predictable timings, presumably due to congestion from other network traffic. The router used for figure \ref{fig:latencytp} was dedicated, no other devices were connected. The effect is clear, with over 80\% of requests taking less than 30ms on the standalone router, vs just 50\% under 30ms on the home router.  The tp-link router was chosen for the solution. An average latency of 30ms gives a refresh rate of over 30Hz, which is sufficient to show no stuttering.
            
\section{Evaluation}

        We have shown that this solution is adequate for correcting large drifts to within 3.3\textdegree{} within only a few seconds of appropriate accelerations, and that the long term accuracy of predictions can be as low as 1.1\textdegree{} RSME under movement of approximately 1 rad\(\cdot \text{s}^{-1}\). These are good results that demonstrate the viability of this algorithm, and IMU tracking in general, for use in microgravity environments.

        \subsection{Strengths and weaknesses of solution}
            The drift correction method makes use of often-discarded centripetal and tangential acceleration data. Making full use of the kinematic chain setup, it extracts maximal information from the system, not discarding data or making assumptions like most IMU systems do. This allows the system to operate in environments that existing solutions cannot: namely, a microgravity space environment without magnetic fields. The system is flexible enough to work in arbitrary gravitational fields, and their presence provides additional accelerations that aid drift correction. The system could therefore operate on other planetary bodies, in free space, or on Earth.
            
            Existing solutions tend to use accelerometers just to estimate the gravity vector. Therefore, any other accelerations degrade performance of the system as it makes the gravity estimate inaccurate. In contrast, this novel approach performs better under fast movements as the accelerations produced give more information about limb orientations, as detailed in results part \ref{accelpredresults}.
            
            For this work we ordered 15 boards, totalling \$85 (£61), including shipping. This price is quite high due to the small batch order. At scale, the unit cost would be less than \$4 (£2.70) per board. Therefore, full body tracking with 15 sensors could be achieved for under £45/person. Taking into account the ESP-8266 hub, battery, and wires we can estimate that full-body tracking with a system like this would cost no more than £55/person at scale. These economics are very favourable for increased adoption of this technology. Such a system is now accessible to general consumers, not just organisations. IMU body tracking is therefore feasible economically for consumer applications in VR and sports analysis.
            
            The effectiveness of the drift correction is dependent on the strength and variability of movement in the kinematic chain for the following reason. With no accelerations, there will be unbounded drift in all axes; with accelerations in one direction, like with gravity, there will be unbounded drift around the axis defined by the acceleration vector (yaw in the Earth case), as seen in figure \ref{fig:correctionexample} before the lateral accelerations begin; however, as the acceleration vector changes direction, drift can be compensated for relative to a new axis. These are the accelerations of each joint, so there needn't be gravity or acceleration at the root of the chain.
            
        \subsection{Limitations}
            It's important to note that while all child orientations are correct relative to the root sensor, there is no way to compensate for drift in the root. This is because there are no absolute references without gravity or magnetism. Of course it couldn't be any other way as ``up" has no meaning without gravity - global coordinates are arbitrary. Nevertheless, it is an inherent limitation that the global orientation of the system cannot be known. The root is like a single IMU, so standard approaches could be used to correct tilt and pitch in a gravitational field, but the yaw of the root will still drift without a magnetometer.
        
            Since the drift correction method relies on local accelerations of varying direction, the solution would perform weakly under little or homogeneous accelerations. For a kinematic chain that moves very slowly, the dead-reckoning drift may overpower any correction efforts. However, as soon as movement resumed the system would begin to correct.
            
            The mathematics of this solution need to be generalised more to work effectively on a human body. There is currently no distinction between the sensor and the limb frame, so mounting on a body would be difficult due to muscle and skin creating irregular orientations. This could be generalised out of the equations though, so it's not an inherent limitation.
            
            The current network design will not scale to full body tracking, as 15 sensors could cycle at at most 13Hz (table \ref{correctionrate}). Furthermore, the WiFi communication may suffer due to currently performing separate HTTP requests each update instead of streaming the data.
        \subsection{Comparison with state-of-the-art}
        
        The XSens Awinda \cite{xsens-mtw-awinda} works for full body tracking and is completely wireless. In contrast this solution will not scale in its current form and is wired. However, this is just a prototype and these limitations could be overcome in a new iteration. Their system can correct for drift in all axes, including the root of the kinematic chain, unlike this system which can not correct in the root. 
        
        The XSens Awinda synchronises to within 10$\mu$s, whereas this solution works over I\textsuperscript{2}C so has higher delays: the acceleration given to a child sensor will be approximately 5ms outdated. Our stated latency is lower for up to 3 sensors, though quickly becomes worse with more sensors. However, these problems could be overcome through better time optimisation and pipe-lining.  They state a correction accuracy of to within  0.75\textdegree{} RMSE for roll and pitch, and 1.5\textdegree{} RMSE for yaw, which is better than our approximate 3.3\textdegree{} for quick correction. Further experimentation is required to determine how quickly the system stabilises in the long-term, but the short-term variations of our drift correction approach were found to be 1.1\textdegree{} RMSE under movement, so with further refinement an accuracy approaching the mtw-Awinda may be achievable.
        
        Unlike any existing solutions \cite{survey-IMU}, XSens included, the contribution in this paper can function in microgravity without reliance on magnetic fields. This is, to the best of our knowledge, the only IMU body tracking system that could function in space, for example on the ISS. We also demonstrated the effectiveness of correlating rotational kinematics with linear accelerations \cite{body-constraints}, showing it to be 4.9x more effective than not doing it for fast movements.
        
        \subsection{Project appraisal}
            
            This project was self-proposed and changed direction significantly over time. Work remained well organised throughout these changes, and the end result marks a strong achievement. A risk was taken in designing a novel algorithm that may not work well, but it was found to give good results and has potential to be a meaningful contribution to the literature.
            
            This project was an ambitious undertaking which tied together a variety of technical domains. Electronic engineering was required when working with the hardware, software engineering for the implementations, and mathematics and physics were required for the circular motion and quaternion mathematics of the algorithm. Much of this learning had to be done specifically for the project. Despite the large scope, the work remained focused and ultimately achieved what it set out to do.
            
            Time could have been managed better, to allow for additional testing and refinement, as well as mounting the sensors on a human. For organic body mounting, however, 3D printed casings would have been required for the sensors, and due to COVID-19 preventing travel and lab access this was not possible.

\section{Conclusions}

    We set out to demonstrate the viability of IMU kinematic chain tracking in novel environments, namely microgravity and in the absence of magnetic fields. We have provided an outline of the topic area and explored the state of the literature and current commercial solutions.
    
    The novel algorithm presented in the work is capable of correcting for drift in gyroscope-based dead-reckoning in the absence of gravitational fields and magnetic fields. It does this utilising just local accelerations at each joint to correct drift in the limbs' orientations. The root of the kinematic chain cannot be drift corrected, but all child limbs can, relative to the root.
    
    We had custom PCBs manufactured and implemented a body-sensor-network to allow for inter-sensor communication. The algorithm is run on each sensor and the data is made available over WiFi, potentially even over the internet, to any client software. Various visualisations of the system were developed in Blender to show the system working.
    
    We found that the novel method for drift correction can correct yaw drift with just acceleration data, and that the method would be able to correct drift even in the absence of gravity. Overall the project was a success that has pushed the state-of-the-art, demonstrating the viability of IMU body tracking in the challenging environments that space presents.

    This solution is still in the early prototype stage, so there are several suggestions for further work to be done. The implementation has room for improvement through optimisation of the architecture. Pipe-lining could be added to the inter-sensor communication to decrease latency and improve synchronisation. The WiFi communication protocol used could also be switched from single-packet HTTP requests to TCP or even UDP streaming.
    
    The algorithm could be generalised to make a distinction between the sensor and limb frame, making human body placement more viable. The system could also be tested on a larger robotic kinematic chain with more segments, as the system here was only evaluated on the 2 segment boom arm.

    Refinement to the filters could give improved accuracy. More options could be investigated for functions $\beta$ and $\Gamma$, as only basic options were used in this implementation. An additional measure could be employed to weight sensitivity to accelerations along different axes. For example, if acceleration has been present in one direction for a long time, tangential accelerations will contain a lot more information than the existing one. 

    Testing in a microgravity environment would be required to fully evaluate a system like this. Zero-g aeroplane services exist that could be used if testing on a space station isn't viable.
    
    \vspace{\baselineskip}
    
    It is reassuring to know that even well-developed technologies can be innovated upon in the increasing drive to move out into space.

\bibliography{projectpaper}

\end{document}